\documentclass[aps,twocolumn,prd,showpacs,floatfix]{revtex4-1}

\usepackage{bm}
\usepackage{amsmath}
\usepackage{nicefrac}
\usepackage{amsthm}
\usepackage{amssymb}
\usepackage{graphicx}

\newcommand{\eref}[1]{(\ref{#1})}
\newcommand{\eps }{\varepsilon }

\begin{document}

%%%%%%%%%%%%%%%%%%%%%%%%%%%%%%%%%%%%%%%%%%%%%%%%%%%
\title{Dense spectrum of resonances and spin-$1/2$ particle capture in a
near-black-hole metric}
\author{G. H. Gossel}
\affiliation{School of Physics, University of New South Wales, Sydney 2052,
Australia}
\author{J. C. Berengut}
\affiliation{School of Physics, University of New South Wales, Sydney 2052,
Australia}
\author{V. V. Flambaum}
\affiliation{School of Physics, University of New South Wales, Sydney 2052,
Australia}
\author{G. F. Gribakin}
\affiliation{School of Mathematics and Physics,
Queen's University, Belfast BT7 1NN, Northern Ireland, UK}

\date{\today}

\begin{abstract}
We show that a spin-1/2 particle in the gravitational field of a massive body of radius $R$ which slightly exceeds the Schwarzschild radius $r_s$, possesses a dense spectrum of narrow resonances. Their lifetimes and density tend to infinity in the limit $R \to r_s$. We determine the cross section of the particle capture into these resonances and show that it is equal to the spin-1/2 absorption cross section for a Schwarzschild black hole. Thus black-hole properties may emerge in a non-singular static metric prior to the formation of a black hole.
\end{abstract}


\maketitle  
%%%%%%%%%%%%%%%%%%%%%%%%%%%%%%%%%%%%%%%%%%%%%%%%%%%
\section{Introduction}
In this work we consider scattering of massless spin-1/2 particles by the
gravitational field of finite-sized bodies whose radius $R$ slightly exceeds
the Schwarzschild radius $r_s = 2GM/c^2$. Here $M$ is the mass of the body,
$G$ is the universal gravitation constant, and $c$ is the speed of light.
The spacetime around the body is described using a suitable metric to model
its interior, which is joined at the surface of the body to the standard
Schwarzschild metric outside. We find that for $R$ approaching $r_s$, the
scattering is characterized by a dense spectrum of narrow resonances, i.e.,
metastable states whose lifetime and energy density tend to infinity in the
black-hole limit. A particle that enters such states is trapped on the
interior of the body for a time $\tau \sim \hbar/\Gamma_n$, where
$\Gamma_n$ is the width of a given resonance.

For $R \to r_s$ both the energy spacing $D$ between the resonances and their
width $\Gamma_n$ tend to zero, and the lifetime $\tau \rightarrow \infty$.
At the same time the ratio $\Gamma _n/D$ for the fixed energy of the particle
remains finite. This allows one to define the cross section for particle
capture into these long-lived states using the optical model \cite{LLV3},
i.e., averaging over a small energy interval containing many resonances. In
doing so we recover the low-energy limit of the absorption cross section for a
pure black hole (with the boundary condition of complete absorption at the
event horizon) derived by Unruh: $\sigma = \frac{1}{4} \pi r_s^2$ for the
$s_{1/2}$ and $p_{1/2}$ partial waves \cite{Unruh}. The total absorption
cross section considering all partial-wave contributions is
$\frac{1}{2}\pi r_{s}^{2}$, as only the $s_{1/2}$ and $p_{1/2}$
waves have nonvanishing cross sections at zero energy. Thus
we observe that at low incident particle energies the absorption properties of
a body with $R>r_s$ resemble those of a black hole. 

It is worth noting that possible inelastic processes, such as radiation by the
particles captured in the long-lived resonances, do not change $\sigma_a$.
The presence of inelastic processes increases the total width of the resonances,
$\Gamma_{\mathrm{tot}}=\Gamma_n+\Gamma_{\mathrm{inel}}$, but this quantity drops
out of the energy averaged (optical) total capture cross section $\sigma _a$, leaving only the
dependence on the elastic width $\Gamma_n$ \cite{LLV3}.

As in the previous work on the scalar (spin-0) case \cite{Flambaum2012}, our
calculations are performed twice: numerically (without approximations) and using analytical
approximations, with good agreement between the two. In contrast to the spin-0
case we consider scattering for arbitrary angular momenta.

This work is closely related to the case of massive spin-1/2 particles trapped by a near-black-hole gravitational field, considered in \citep{SpencerBound}. The authors showed that the bound-state energy spectrum collapses and becomes quasi-continuous in the black-hole limit. The collapse of the positive-energy resonance spectrum in the black-hole limit found in this work shows similar behaviour.
%%%%%%%%%%%%%%%%%%%%%%%%%%%%%%%%%%%%%%%%%%%%%%%%%%%%%%%%%%%%%%%%%%%
\section{Dirac equation in curved spacetime}
\subsection{Radial equation}

Consider a curved spacetime with the static, spherically symmetric metric:
\begin{equation}\label{eq:metric}
ds^2 = a(r) dt^2-b(r)dr^2 -r^2 d\Omega^2,
\end{equation}
where $a(r)$ and $b(r)$ are positive functions. The Dirac equation for a massless spin-1/2 particle in the above metric may be represented as two coupled equations (derived in \citep{SpencerBound}) for upper and lower components of the wavefunction $f(r)$ and $g(r)$, given by
\begin{gather}
\frac{df({r})}{dr} +\sqrt{b(r)}\frac{\kappa}{r} f({r}) - \eps \sqrt{\frac{b(r)}{a(r)}}g({r}) = 0,\notag{} \\
\frac{dg({r})}{dr} -\sqrt{b(r)} \frac{\kappa}{r} g({r}) + \eps \sqrt{\frac{b(r)}{a(r)}} f({r}) = 0.
\label{eq:RadialDirac}
\end{gather}
Here $\kappa =\mp (j+\frac{1}{2})$, where
$j=l\pm \frac{1}{2}$ is the total angular momentum and $l$ is the orbital
angular momentum. The equations \eref{eq:RadialDirac} can be recast as the following second-order differential equation for $f(r)$ in a given partial wave
\begin{align}\label{eq:GenEqn}
&f''(r)+\frac{b(r)}{2a(r)}\left[\frac{a(r)}{b(r)}
\right]'f'(r)\\ \notag{}
&+\left\{\frac{\varepsilon^2 b(r)}{a(r)}+\frac{\kappa \sqrt{b(r)}}{r^2}
\left[\frac{r a'(r)}{2a(r)}-\kappa \sqrt{b(r)}-1\right]\right\}f(r) = 0.
\end{align}

\subsection{Interior solution}
\label{sec:Interior}
Changing the radial variable to to the Regge-Wheeler ``tortoise'' coordinate
$r^*$ defined by $d r^*=\sqrt{b(r)/a(r)}\,d r$, we can transform
Eq.~\eref{eq:GenEqn} to the following Schr\"odinger-like equation for $f(r^*)$,
\begin{equation}\label{eq:Sch}
\frac{d^2 f}{d{r^*}^2}+\left\{\eps ^2+\frac{\kappa r}{2\sqrt{b(r)}}
\left[\frac{a(r)}{r^2}\right]'-\frac{\kappa^2 a(r)}{r^2}\right\}f=0.
\end{equation}

The metric outside of a massive nonrotating spherical body is given by
the Schwarzschild solution,
\begin{equation}\label{eq:Schwarz}
a(r)= 1-r_s/r,\quad b(r)=(1-r_s/r)^{-1}.
\end{equation}
On the surface of the body $a(R)=1-r_r/R$. Since the metric is
continuous, smooth and monotonic, then in the near-black-hole limit
($r_s\rightarrow R$) the interior metric $a(r)\to 0$ for all
$0\leq r\leq R$, as the time slows down inside the gravitational potential. In this regime
the first term in brackets in Eq.~(\ref{eq:Sch}) dominates for all
except very small energies, and for all distances, except near the origin.
This means that the solution away from the origin describes free motion
in the tortoise coordinate,
\begin{equation}\label{eq:free}
f \simeq \sin (\eps r^* +\phi ),
\end{equation}
where the phase $\phi $ is determined by behaviour of the wave function 
near the origin.

In the vicinity of $r=0$, the dominant coefficient of $f(r^*)$ in
Eq.~(\ref{eq:Sch}) is given by the $r^{-2}$ centrifugal terms. For the specific
metrics we consider (see Sec. \ref{sec:Specifics}), for $r\rightarrow 0$,
$a(r)\simeq a(0) >0$ and
$b(r) \simeq 1$. This also applies to a wider class of static solutions
where the potential is harmonic near the origin. In this case, the centrifugal
term in Eq.~(\ref{eq:Sch}) is $\kappa (\kappa +1)/{r^*}^2=l(l+1)/{r^*}^2$, and
the corresponding phase shift is given by $\phi =-l \pi/2$ \cite{LLV3}.
Hence, away from the origin the interior solution that is regular at the
origin, is given by
\begin{equation}\label{eq:sol_int}
f(r)= \sin \left(\eps \int _0^r \eta (r')dr'-\frac{l\pi }{2}\right),
\end{equation}
where $\eta (r)=\sqrt{b(r)/a(r)}$. [Analysis of Eqs.~(58) and
(59) in \citep{SpencerBound} shows that for the regular solution, $f\propto r^{l+1}$ and
$g\propto r^{l'+1}$, which explains why the phase in Eq.~(\ref{eq:sol_int})
contains $l$ rather than $\kappa $.]

In what follows we solve the scattering problem by matching the logarithmic
derivative of the exterior solution at $r=R$ to the logarithimic derivative
of the interior solution, Eq.~(\ref{eq:sol_int}):
\begin{equation}\label{eq:IntLD}
\left.\frac{f'(r)}{f(r)}\right|_R = \frac{\eps R}{R-r_s}
\cot \left(\eps \Lambda(R) -\frac{l\pi }{2}\right),
\end{equation}
where
\begin{equation}\label{eq:Lambda}
\Lambda(R) = \int_{0}^{R} \eta(r)dr. 
\end{equation}
In Eq.~(\ref{eq:IntLD}) we also used the fact that
\begin{equation}\label{eq:PatR}
\eta (R)=R/(R-r_s),
\end{equation}
since the interior metric matches Eq.~(\ref{eq:Schwarz}) at $r=R$.

Note that in the black-hole limit ($R\rightarrow r_s$) the function
$\Lambda(R)$ tends to infinity. This means that for a fixed energy
$\varepsilon $, the phase of the interior wave function (\ref{eq:sol_int})
is large and the wave function oscillates rapidly, in close analogy with
the case of massless scalar particles with $l=0$ \cite{Flambaum2012}.
%%%%%%%%%%%%%%%%%%%%%%%%%%%%%%%%%%%%%%%%%%%%%%%%%%%%%%%%%%%%%%%%%%%
\subsection{Exterior solution}
Using the exterior Schwarzschild metric, Eq.~(\ref{eq:Schwarz}), in the radial
wave equation (\ref{eq:GenEqn}), we obtain for $r>R$:
\begin{align}\label{eq:ExteriorWave}
&f''(r) + \left(\frac{1}{r-r_s}-\frac{1}{r}\right)f'(r)  \\ \notag{}
&+\left[\frac{\varepsilon^2 r^2}{(r-r_s)^2} +
\frac{\kappa (3r_s-2r)}{2r^{3/2} (r-r_s)^{3/2}} -
\frac{\kappa^2}{r (r-r_s)}\right]f(r) = 0.
\end{align}
%%%%%%%%%%%%%%%%%%%%%%%%%%%%%%%%%%%%%%%%%%%%%%%%%%%%%%%%%%%%%%%%%%%
\subsubsection{Region I ($r \approx R$)}
For near-black-hole metrics ($R\approx r_s$) we can keep only the most singular
terms in the wave equation (\ref{eq:ExteriorWave}) near the boundary
($r\approx R$), defined as Region I. Neglecting less singular terms and
setting $r=r_s$ elsewhere, we have
\begin{equation*}\label{eq:weqRI}
f''+\frac{1}{r-r_s}f' +\left[\frac{\varepsilon^2 r_s^2}{(r-r_s)^2} +
\frac{\kappa }{2r_s^{1/2} (r-r_s)^{3/2}}\right]f = 0.
\end{equation*}
The exact solution of this equation are the Bessel
functions $J_{4i\eps r_s}(\rho )$ and $Y_{4i\eps r_s}(\rho )$, where
$\rho = \sqrt{8\kappa}\sqrt[4]{(r-r_s)/r_s}$. At low energies
$\varepsilon r_s\ll 1$, using the lowest-order terms in the expansion of
the Bessel functions, gives the wave function in Region I as
\begin{equation}
\label{eq:ExtSol1}
f_1(r)= \alpha_1 + \beta_1 \ln\left(\frac{r-r_s}{r_s}\right),
\end{equation}
where the constants $\alpha_1$ and $\beta_1$ are determined by matching with
the interior solution at the boundary. 

%%%%%%%%%%%%%%%%%%%%%%%%%%%%%%%%%%%%%%%%%%%%%%%%%%%%%%%%%%%%%%%%%%%
\subsubsection{Region II ($r \gg R$)}
In this region the wave equation (\ref{eq:ExteriorWave}) takes the form
of the nonrelativistic Schr\"odinger equation for a particle with momentum 
$\eps $, angular momentum $l$ and unit mass in the attractive Coulomb field $Z/r$
with the charge $Z=-\eps ^2r_s$. The exterior solution is thus a linear
combination of the regular and irregular Coulomb functions,
\begin{equation}
\label{eq:LargeRSol}
f_2(r) = \alpha_2 F_l (\eps r) + \beta_2 G_l(\eps r),
\end{equation}
which behave asymptotically as $F_l\sim \sin z$ and $G_l\sim \cos z$, where
$z= \varepsilon r+\varepsilon r_s\ln 2\varepsilon r -l\pi/2+\delta_l^C$
and $\delta_l^{C}= \arg [\Gamma (l + 1 - i\varepsilon r_s)]$ is the Coulomb
phase shift.

Following Unruh's matching procedure \cite{Unruh}, we find the
relationships between the coefficients in regions I and II,
for $\kappa<0$ \cite{UnruhNote},
\begin{align}\label{eq:a12-}
\alpha_2 &=\frac{\alpha_1 }{C_{l}(\eps)}
\left(\frac{4}{\varepsilon r_s}\right)^{|\kappa|},\\ \label{eq:b12-}
\beta_2 & = -\frac{\beta_1 C_{l}(\eps) }{4}
\left( \frac{\varepsilon r_s}{4}\right)^{|\kappa|-1} ,
\end{align}
and for $\kappa>0$,
\begin{align}\label{eq:a12+}
\alpha_2 & =\frac{\beta_1 }{4C_{l}(\eps) (2\kappa+1)}
\left(\frac{4}{\varepsilon r_s }\right)^{\kappa+1}, \\ \label{eq:b12+}
\beta_2 &= \alpha _1 C_{l}(\eps) (2\kappa+1)
\left(\frac{\varepsilon r_s }{4}\right)^\kappa .
\end{align}
In these equations $C_{l}(\varepsilon)$ is the Coulomb factor,
\begin{equation}
C_{l}(\varepsilon) = 2^l e^{\pi \varepsilon r_s /2}
\frac{|\Gamma(l+1-i\varepsilon r_s)|}{(2l+1)!}.
\end{equation}
For $\varepsilon r_s\ll 1$ this factor is a constant,
$C_{l}\simeq 1/(2l+1)!!$.
%%%%%%%%%%%%%%%%%%%%%%%%%%%%%%%%%%%%%%%%%%%%%%%%%%%%%%%%%%%%%%%%%%%
\section{$S$-matrix and resonances}
The solution to Eq.~\eref{eq:ExteriorWave} at large distances can be written
as
\begin{equation}\label{eq:fAB}
f(r) \sim  A e^{i z} +B e^{-iz}, 
\end{equation}
The ratio of the coefficients in front of the outgoing and incoming waves
defines the $S$-matrix,
\begin{equation}
S_\kappa = -\frac{A}{B}\exp\left(2i\delta_{l}^{C}\right),
\end{equation}
and the short-range phase shift $\delta $, via $e^{2i\delta }\equiv -A/B$.
Comparing Eq.~\eref{eq:LargeRSol} with \eref{eq:fAB}, we obtain
\begin{equation}\label{eq:AB}
A= \frac{\beta_{2} - i \alpha_{2}}{2},\quad
B= \frac{\beta_{2} + i \alpha_{2}}{2},
\end{equation}
which yields
\begin{equation}\label{eq:S}
S_{\kappa} = -\frac{1-i \alpha_2/\beta_2}{1+i \alpha_2/\beta_2}
\exp\left(2i\delta_{l}^{C}\right).
\end{equation}
Using the values of $\alpha_2$ and $\beta_2$ determined previously we have
\begin{align}\label{eq:rat-}
\frac{\alpha_2}{\beta_2}&=
-\frac{4}{C_l^2}
\left(\frac{4}{\varepsilon r_s}\right)^{2|\kappa |-1}\frac{\alpha_1}{\beta_1}
\quad &(\kappa <0),\\ \label{eq:rat+}
\frac{\alpha_2}{\beta_2}&=\frac{1}{4C_l^2 (2\kappa +1)^2}
\left(\frac{4}{\varepsilon r_s}\right)^{2\kappa +1}\frac{\beta_1}{\alpha_1}
\quad &(\kappa >0).
\end{align}
The elastic scattering cross section is proportional to $|1-S_\kappa |^2$. 
When the $S$-matrix varies rapidly as a function of energy, the cross
section displays resonance maxima for $S_\kappa \approx -1$. At low energies,
$\varepsilon r_s\ll 1$, the Coulomb phase shift is small (and it varies slowly
with energy), and the resonances occur for $\alpha _2/\beta _2=0$
[see Eq.~(\ref{eq:S})]. This corresponds to
\begin{equation}\label{eq:WeakResCondition}
\frac{\alpha_1}{\beta_1}=0 \quad (\kappa<0),\quad 
\frac{\beta_1}{\alpha_1} = 0\quad (\kappa>0).
\end{equation}
These ratios are determined by matching the solution for the Schwarzschild
exterior metric with the interior solution.
%%%%%%%%%%%%%%%%%%%%%%%%%%%%%%%%%%%%%%%%%%%%%%%%%%%%%%%%%%%%%%%%%%%
\subsection{Resonance energies}
Matching the logarithmic derivative of the exterior solution (\ref{eq:ExtSol1})
at $r=R$ to that from Eq.~(\ref{eq:IntLD}) yields
\begin{equation}
\label{eq:AlphaOnBeta}
\frac{\alpha_1}{\beta_1}=\frac{\tan\left[\varepsilon \Lambda (R) -
l\pi /2\right]}{ \eps R}-\ln\left(\frac{R-r_s}{R}\right).
\end{equation}

The resonance conditions, Eq.~\eref{eq:WeakResCondition}, translate into the
tangent function tending to either zero or infinity depending on the sign of
$\kappa $. (There is a small offset due to the logarithmic term but this is
negligible in the $r_s\rightarrow R$ limit, since $\Lambda (R)$ increases much
faster.) Hence, we find the expression for the resonance energies
\begin{equation}\label{eq:AnalyticEn}
\varepsilon_n =\frac{\pi [n +(|\kappa|-1)/2]}{\Lambda(R)},
\end{equation}
which is valid for both $\kappa<0$ and $\kappa >0$.
%Thus the choice of $n-1/2$ in the quantization condition for $\kappa>0$ ensures that $\kappa = -1$ and $\kappa=1$ resonance energies match.% (as the numerics indicates they do and as the Unruh result suggests they should).

Since $\Lambda(R)\rightarrow \infty $ for $R\rightarrow r_s$, the energies of
all resonances tend to zero in the black-hole limit, and the resonance spectrum
``collapses'', as its density tends to infinity. A similar collapse of spectrum is also seen in the bound state case detailed in \citep{SpencerBound}.

%The constant $\zeta$ is due to a phase shift associated with matching the wavefunction at the classical turning point, prior to which there is suppression from the centrifugal barrier. This shift is found to be approximately $\zeta = -l \pi/2$. Thus not only does the centrifugal barrier lower the total phase at infinity by $l \pi/2$, but the position of the first resonance is also shifted giving rise to an additional shift of the same size (when approximating the phase by a linear function).
%%%%%%%%%%%%%%%%%%%%%%%%%%%%%%%%%%%%%%%%%%%%%%%%%%%%%%%%%%%%%%%%%%%
\subsection{Resonance widths}
The full resonance condition states that resonances correspond to poles of the $S$-matrix
at energies $\varepsilon = \varepsilon_n -i\Gamma _n/2$ which lie below
the real axis in the complex-energy plane \cite{LLV3}. According to
Eq.~(\ref{eq:S}), this occurs when
\begin{equation}
\label{eq:FullResC}
1+\frac{i \alpha_2}{\beta_2} = 0.
\end{equation}
To determine the resonance widths $\Gamma _n$, we use Eqs. (\ref{eq:rat-}) and
(\ref{eq:rat+}) and expand the ratio
$\alpha _1/\beta _1$ (for $\kappa <0$) or $\beta _1/\alpha _1$ (for $\kappa >0$)
to first order about the resonance energy $\varepsilon _n$.
For example, for negative $\kappa $ we use
\begin{equation}
\frac{\alpha_1}{\beta_1} \simeq
\left(\frac{\alpha_1}{\beta_1}\right)'(\varepsilon-\varepsilon_n)
= \left(\frac{\alpha_1}{\beta_1}\right)' (-i\Gamma_n/2),
\end{equation}
thus Eqn.~\eref{eq:FullResC} may be written (for negative $\kappa$) as
\begin{equation}
1+f(\eps) \frac{\Gamma_n}{2} \left(\frac{\alpha_1}{\beta_1}\right)'=0,
\end{equation}
where the prime denotes differentiation with respect to $\varepsilon $,
the derivative is evaluated using Eq.~(\ref{eq:AlphaOnBeta}) at the point
where $\tan [\varepsilon \Lambda (R) - l\pi /2]=0$, and $f(\eps)$ is defined by Eqn.~\eref{eq:rat-}. Hence, we obtain for $\kappa <0$,
\begin{equation}
\label{eq:Gam-}
\Gamma_n  =\frac{2C_l^2R}{\Lambda(R) r_s}\left(\frac{\varepsilon r_s}{4}\right)^{2|\kappa|},
\end{equation}
and following a similar procedure expanding $\beta_1/\alpha_1$ for $\kappa >0$,
\begin{equation}
\label{eq:Gam+}
\Gamma_n =\frac{2C_l^2(2\kappa+1)^2r_s}{\Lambda (R)R}\left(\frac{\varepsilon r_s}{4}\right)^{2\kappa}.
\end{equation}
The widths must be evaluated at $\varepsilon =\varepsilon _n$ from
Eq.~(\ref{eq:AnalyticEn}).
%%%%%%%%%%%%%%%%%%%%%%%%%%%%%%%%%%%%%%%%%%%%%%%%%%%%%%%%%%%%%%%%%%%
\subsection{Cross sections}

Comparing the above expressions for the widths with the resonance energy
spacing $D=\eps_{n+1}-\eps_n=\pi /\Lambda (R)$ [see Eq.~(\ref{eq:AnalyticEn})], 
we see that $\Gamma _n\ll D$ at low energies, i.e., $\varepsilon r_s\ll 1$.
In this case one can consider the cross section of capture into the resonances.
This cross section corresponds to the optical-model energy-averaged
\textit{absorption cross section} \cite{LLV3}, which is given by
\begin{equation}
\sigma_\kappa^{(a)} = |\kappa |\frac{2\pi^2 \Gamma_n}{\varepsilon^2 D},
\end{equation}
for a particular partial wave $\kappa $.

Using Eqs.~(\ref{eq:Gam-}) and (\ref{eq:Gam+}), we obtain the resonant
absorption cross sections for the near-black-hole metric:
\begin{equation}\label{eq:sig}
\sigma_\kappa^{(a)}=
\begin{cases}
\dfrac{1}{4}\pi r_s^2|\kappa |C_l^2
\left(\dfrac{\varepsilon r_s}{4}\right)^{2|\kappa |-2}
&\!\!(\kappa <0),\\[9pt]
\dfrac{1}{4}\pi r_s^2|\kappa |C_l^2(2\kappa +1)^2
\left(\dfrac{\varepsilon r_s}{4}\right)^{2\kappa -2}
&\!\!(\kappa >0).
\end{cases}
\end{equation}
Note that, unlike the resonance energies and widths, the resonance capture
cross sections do not contain $\Lambda (R)$, and hence, are \emph{independent}
of the interior metric used, as long as the metric satisfies the assumptions made in Sec.~\eref{sec:Interior}.

In the zero-energy limit only two partial waves give nonzero contributions,
namely, $s_{1/2}$ and $p_{1/2}$ ($|\kappa| = 1$). The corresponding
cross sections $\sigma _{\mp 1}^{(a)}=\frac{1}{4}\pi r_s^2$ are in agreement
with Unruh's result $\sigma_\mathrm{tot}^{(a)}=\sum\limits_{\kappa} \sigma_{\kappa}^{(a)}=\frac{1}{2}\pi r_{s}^{2}$ \cite{Unruh}.

%%%%%%%%%%%%%%%%%%%%%%%%%%%%%%%%%%%%%%%
\section{Numerical results for specific interior metrics}\label{sec:Specifics}

In this section we present calculations involving two specific interior metrics that allow the $r_s \rightarrow R$ limit to be taken: the Florides \cite{Florides74} and Soffel \cite{Soffel77} metrics. Specifically, we verify the analytics provided previously with numerically calculated resonance widths and energies via the short range phase shift. To calculate this short range phase $\delta_\textrm{I}$ we solve the second-order differential equation (\ref{eq:GenEqn}) numerically, for given $a(r)$ and $b(r)$, with the boundary condition $f(0)=r^{l+1}$, $f '(0)=(l+1)r^l$ using {\em Mathematica} \cite{math}. This solution provides a real boundary condition for the {\em exterior} wave function at $r=R$. (We set $R=1$ in the numerical calculations). Equation \eref{eq:ExteriorWave} is then integrated outwards to large distances $r\gg r_s$. In this region Eq.~(\ref{eq:ExteriorWave}) takes the form of a nonrelativistic Shr\"odinger equation for a particle with momentum $\eps $ and unit mass in the Coulomb potential with charge $Z=-r_s\eps ^2$. Hence, we match the solution with the asymptotic form \cite{LLV3}
\begin{equation}\label{eq:CoulombMatch}
f(r)\propto \sin [\varepsilon r - (Z/\varepsilon ) \ln 2 \varepsilon r +\delta_C+ \delta-l \pi/2 ]
\end{equation}
where $\delta_C = \arg \Gamma (1+l+i Z/\varepsilon ) $ is the Coulomb phase shift,
% $\Gamma (x)$ being the Euler gamma function,
and determine the short-range phase shift $\delta $.

The numeric widths and positions of the resonances are then extracted from this phase shift by fitting it to the Breit-Wigner profile
\begin{equation}
\delta(\eps )= A+\arctan\left[\frac{\eps-\eps_n}{\Gamma_n/2}\right]
\end{equation}
in the region of an isolated resonance ($\eps\approx \eps_n$), where $A$ is a constant offset.
\begin{figure}[tb]
\begin{center}
\includegraphics*[width=0.48\textwidth]{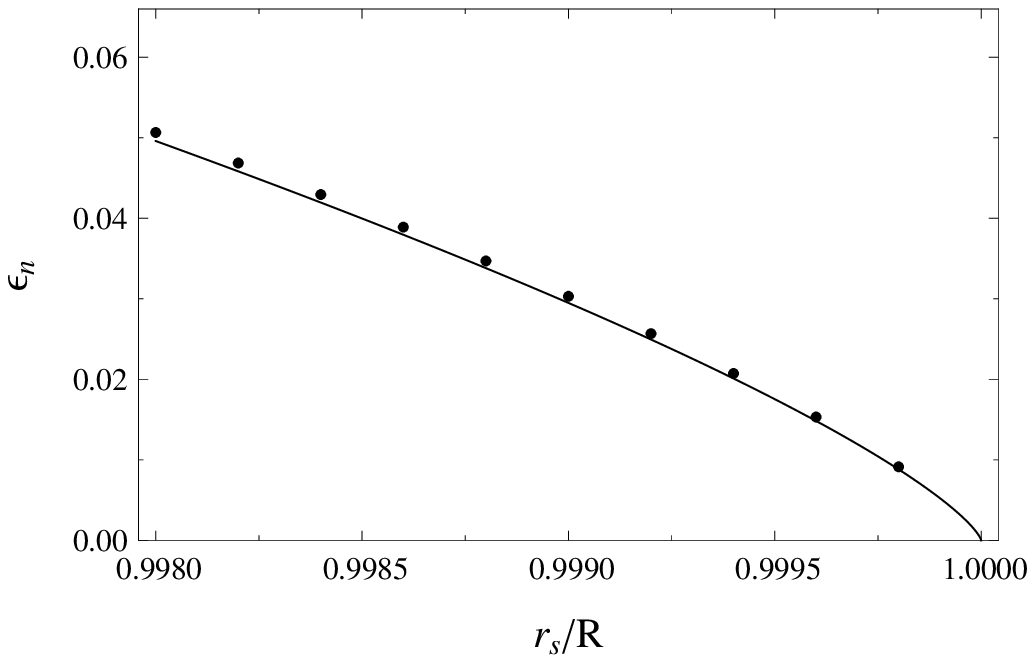}
\caption{Energies of the $n=2$ resonance in the Florides metric. Closed circles  indicate numeric data, the solid line indicates analytic $\varepsilon_n$ given by Eqn.~\eref{eq:AnalyticEn} with $\lambda$ given by Eqn.~\eref{eq:FloridesL}.
\label{fig:FloridesEn}}
\end{center}
\end{figure}
\begin{figure}[tb]
\begin{center}
\includegraphics*[width=0.48\textwidth]{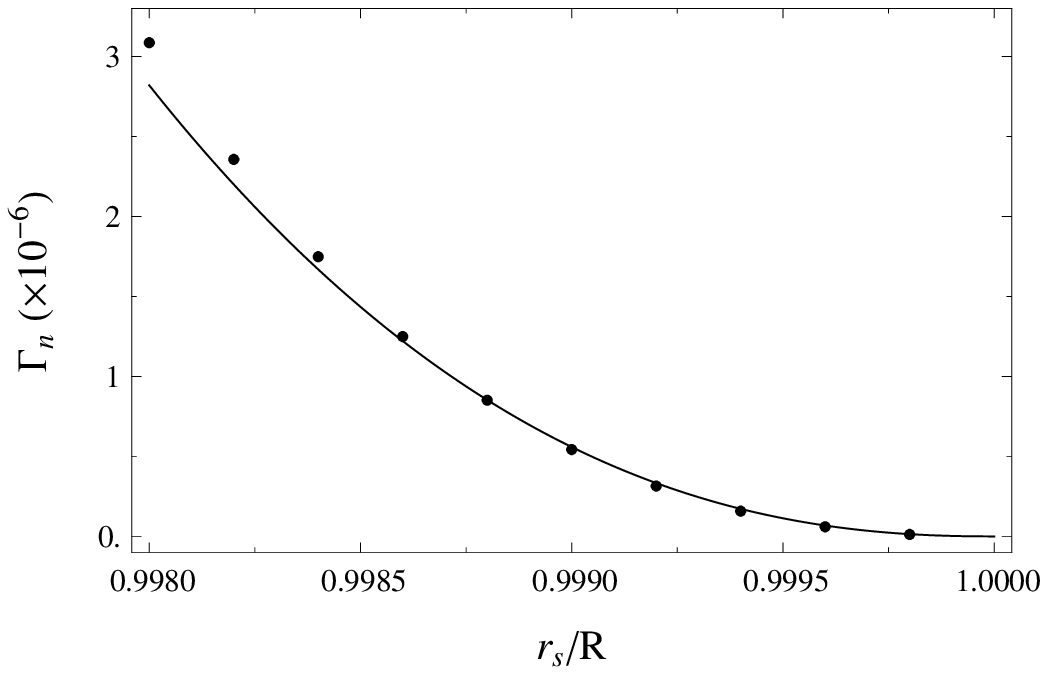}
\caption{Width of the $n=2$ resonance in the Florides metric. Closed circles indicate numeric data, the solid line indicates analytic $\Gamma_n$ given by Eqn.~\eref{eq:Gam-} with $\lambda$ given by Eqn.~\eref{eq:FloridesL}.
\label{fig:FloridesW}}
\end{center}
\end{figure}
\subsection{Florides Interior}
The Florides metric is characterized by
\begin{equation}
\label{eq:FInt}
a(r)=\frac{(1-r_s/R)^{3/2}}{\sqrt{1-r_sr^2/R^3}},\quad
b(r)=\left(1-\frac{r_s r^2 }{R^3}\right)^{-1}.
\end{equation}
This in turn leads to 
\begin{align}
\label{eq:FloridesL}
\Lambda(R)_\textrm{F} \overset{r_s \rightarrow R}{=}& \frac{\pi^{3/2} R}{\sqrt{2} \Gamma(\nicefrac{1}{4}) \Gamma(\nicefrac{5}{4}) (1-r_s /R)^{3/4}}\notag{} \\
\approx& 1.198 (R-r_s)^{-3/4}.
\end{align}
The resulting resonance energies and widths are compared with their numeric counterparts in Figures \eref{fig:FloridesEn} and \eref{fig:FloridesW} respectively.
\subsection{Soffel Interior}
\begin{figure}[tb]
\begin{center}
\includegraphics*[width=0.48\textwidth]{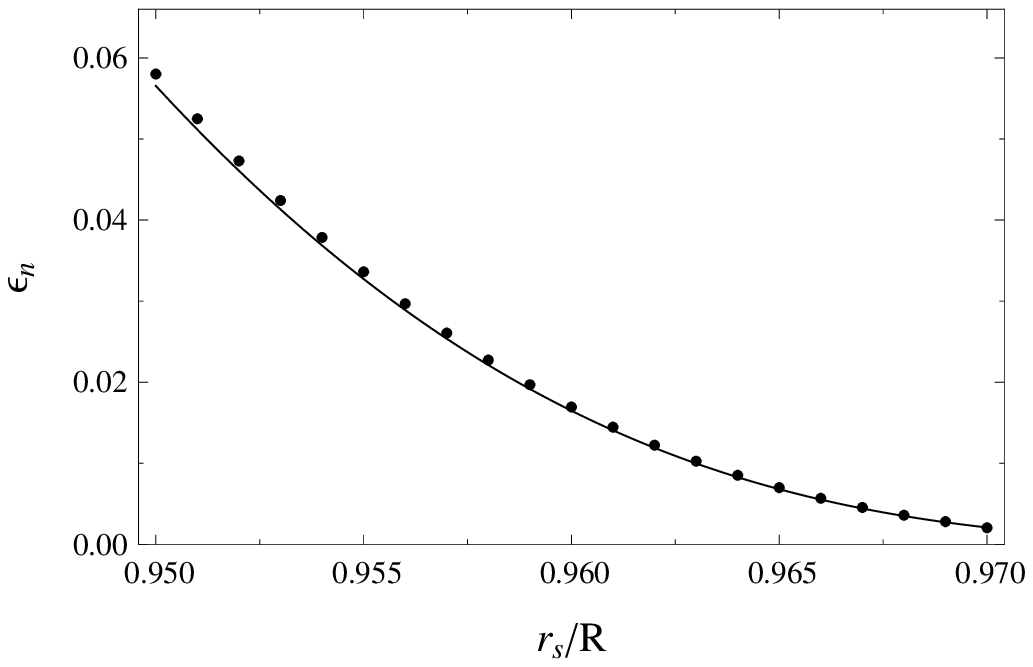}
\caption{Energies of the $n=4$ resonance in the Soffel metric. Closed circles  indicate numeric data, the solid line indicates analytic $\varepsilon_n$ given by Eqn.~\eref{eq:AnalyticEn} with $\lambda$ given by Eqn.~\eref{eq:SoffelL}.
\label{fig:SoffelEn}}
\end{center}
\end{figure}

\begin{figure}[!tb]
\begin{center}
\includegraphics*[width=0.48\textwidth]{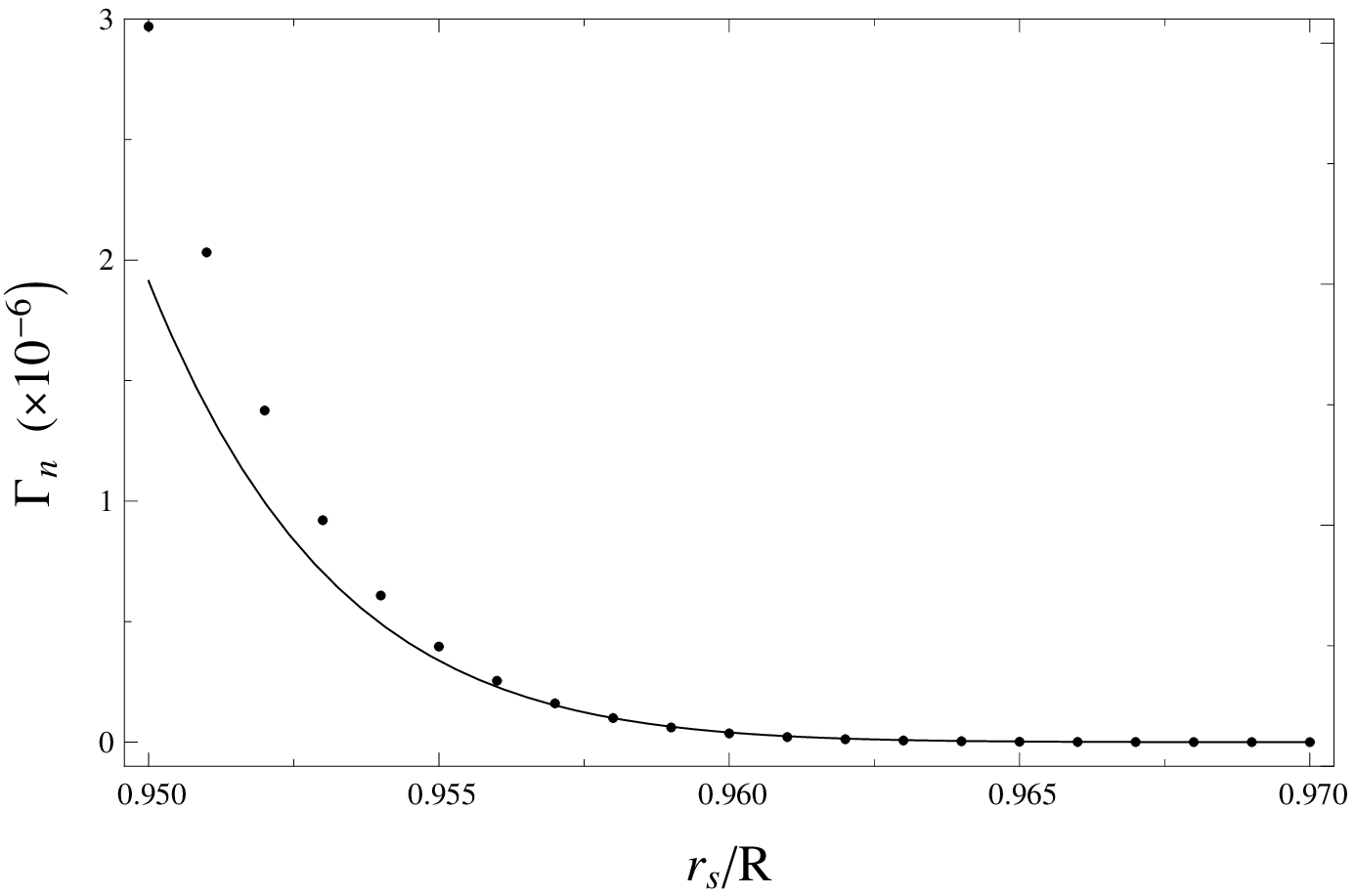}
\caption{Widths of the $n=4$ resonance in the Soffel metric. Closed circles indicate numeric data, the solid line indicates analytic $\Gamma_n$ given by Eqn.~\eref{eq:Gam-}, with $\lambda$ given by Eqn.~\eref{eq:SoffelL}.
\label{fig:SoffelW}}
\end{center}
\end{figure}

The Soffel metric is characterized by \cite{Soffel77}
\begin{equation}
\label{eq:SInt}
a(r)= \left(1-\frac{r_s}{R}\right) 
\exp \left[-\frac{r_s (1-r^2/R^2)}{2R (1-r_s/R)}\right],
\end{equation}
with $b(r)$ equal to that of the Florides case. This in turn leads to 
\begin{align}
\label{eq:SoffelL}
\Lambda(R)_\textrm{So} \overset{r_s \rightarrow R}{=}&R\sqrt{\pi}\exp\left[\frac{r_s/R}{4(1-r_s/R)}\right]\notag{} \\
\end{align}
Analytic and numeric $\varepsilon_n$ and $\Gamma_n$ for the Soffel metric are compared in Figures \eref{fig:SoffelEn} and \eref{fig:SoffelW} respectively.

\section{Conclusions}
%%%%%%%%%%%%%%%%%%%%%%%%%%%%%%%%%%%%%%%
The problem of scattering of low-energy spin-1/2 particles from a massive static spherical body has been considered. We have shown that as in the spin-0 case, approaching the black hole case gives rise to a dense spectrum of long lived resonances. Similar to the scalar case, we show that the existence and structure of these resonances gives rise to effective absorption in the purely potential scattering problem. This allows us to construct an absorption cross section for bodies near the black hole threshold which matches known results for the pure black hole case in the low energy limit.
%%%%%%%%%%%%%%%%%%%%%%%%%%%%%%%%%%%%%%%

\begin{acknowledgments}
We thank A. V. Korol for useful discussions.
\end{acknowledgments}


\begin{thebibliography}{9}
\bibitem{UnruhNote} Note that equation (46) in Unruh's paper \cite{Unruh}
contains a misprint: the sign of the second term should be minus, not
plus. Consequently, the right-hand side of the second equation (51) should also
contain a minus.
\end{thebibliography}
\end{document}